\documentclass[10pt,aps,showpacs,byrevtex]{revtex4}
\usepackage{bm}
\usepackage{amsmath}
\begin{document}
\title{\small KLEIN PARADOX FOR OPTICAL SCATTERING FROM EXCITED
TARGETS}
\author{\scshape{O. Panella}}
\affiliation{Istituto Nazionale di Fisica Nucleare, Sezione di Perugia,
Via A. Pascoli, Perugia Italy}
\author{\scshape{Y.N. Srivastava and A. Widom}}
\affiliation{Physics Department \& INFN, University of Perugia, Perugia
Italy}
\affiliation{Physics Department, Northeastern University, Boston MA
USA}

\date{June 30, 2006 {\sf  Journal-Ref: IJMPA 21, 3279-3288, 2006} }
\preprint{quant-ph/0105010}

\begin{abstract}
The well known Klein paradox for the relativistic Dirac wave equation
consists in the computation of possible ``negative probabilities''
induced
by certain potentials in some regimes of energy. The paradox may be
resolved
employing the notion of electron-positron pair production in which the
number of electrons present in a process can increase. The Klein
paradox
also exists in Maxwell's equations viewed as the wave equation for
photons.
In a medium containing ``inverted energy populations'' of excited
atoms,
e.g. in a LASER medium, one may again compute possible ``negative
probabilities''. The resolution of the electromagnetic Klein paradox is
that when the atoms decay, the final state may contain more photons
then
were contained the initial state. The optical theorem total cross
section
for scattering photons from excited state atoms may then be computed as
negative within a frequency band  with matter induced amplification.
\end{abstract}
\maketitle

\section{Introduction}

At a time in which the properties of relativistic electrons (implicit
in the  Dirac equation\cite{1} for the spinor wave function) were not
so well understood, Klein predicted a ``paradox'' associated with
the problem of reflecting an electron off a step potential.
The reflection coefficient
\begin{math} R \end{math} and the transmission coefficient
\begin{math} P \end{math} (induced by the step potential) obey an
expected sum rule
\begin{equation}
R+P=1\ \ {\rm  (exact)}.
\end{equation}
However, for some step potential heights and for some incident energies
\begin{equation}
R>1\ \ {\rm  (possible)}.
\end{equation}
Eqs.(1) and (2) imply the possibility of a negative transmission
``probability'' coefficient; i.e. \begin{math} P<0 \end{math} is
possible
and therein lies the paradox.

The Dirac wave equation for electrons and positrons allows for a
physical picture which clarifies the meaning of the Klein
paradox\cite{2}
and thus removes it\cite{3,4,5,6} as a possible objection to the Dirac
theory. When an electron hits the step potential, it is possible to
create an electron-positron pair. For such an event, two electrons may
be reflected from the step potential when only one electron was
incident.
The added positron may then be transmitted into the step potential.
Such an event, if sufficiently probable, may yield
\begin{math} R>1 \end{math} since more electrons may be
reflected from the step potential than are incident. Furthermore,
\begin{math} P<0 \end{math} may be interpreted to mean that the current
of the transmitted positron wave is directed oppositely to that
which would have been carried by an electron wave moving at the same
velocity.

Our purpose is to point out that the Klein paradox\cite{7,8,9} is
intrinsic to relativistic particles\cite{10,11,12,13} and in particular
applies to the photon. For the electromagnetic case, the wave equations
for the photon in space-time are merely Maxwell's equations. To
understand the nature of the Klein paradox for photons we may recall
the classical results discussed by Rayleigh. Suppose a spherically
symmetric target with a fluctuating dipole moment which scatters an
electromagnetic wave. We suppose that the electric dipole moment
\begin{math} {\bm p}(t)  \end{math}
response to an applied electric field
\begin{equation}
{\bm E}(t)={\Re}e\left\{{\bm E}_0 e^{-i\zeta t}\right\} \ \ {\rm
with}\ \ {\Im}m (\zeta )>0,
\end{equation}
is given by
\begin{equation}
{\bm p}(t)={\Re}e\left\{\alpha (\zeta ){\bm E}_0 e^{-i\zeta
t}\right\}.
\end{equation}
Here \begin{math} \alpha (\zeta ) \end{math} is the target
polarizability.

Rayleigh asserted that the elastic scattering amplitude for an
incident electromagnetic wave onto the polarizable target is given
in the dipole approximation by
\begin{equation}
F_{i\to f}=\left(\frac{\omega }{c}\right)^2 \, {\bm e}_f^* \cdot {\bm
e}_i \, \alpha (\omega +i0^+)
\end{equation}
where \begin{math} {\bm e}_i  \end{math} and
\begin{math} {\bm e}_f \end{math}
are, respectively, the initial and final polarization vectors.
Averaging over initial polarization and summing over final
polarizations yields the elastic differential cross section
\begin{eqnarray}
\left(\frac{d\sigma_{el}}{d\Omega }\right)&=&
\frac{1}{2}\sum_i \sum_f \left|F_{i\to f}\right|^2 \nonumber \\
 &=& \frac{1}{2}
\left(1+\cos ^2 \theta \right)\left(\frac{\omega }{c}\right)^4
\left|\alpha (\omega +i0^+)\right|^2,
\end{eqnarray}
which implies for the elastic cross section
\begin{math} \sigma_{el}=\int d\sigma_{el} \end{math} that
\begin{equation}
\sigma_{el}(\omega )=
\left(\frac{8\pi }{3}\right)\left(\frac{\omega}{c}\right)^4
\left|\alpha (\omega +i0^+)\right|^2 .
\end{equation}

For understanding the Klein paradox for electromagnetic waves it is
more important to discuss the total cross section which involves
the optical theorem
\begin{equation}
\sigma_{tot}=\left(\frac{4\pi c}{\omega }\right){\Im}m \big(F_{i\to
i}\big)
\end{equation}
in the form
\begin{equation}
\sigma_{tot}(\omega )= \left(\frac{4\pi \omega }{c}\right)\, {\Im
}m\left[\alpha (\omega +i0^+)\right].
\end{equation}
The expected sum rule is that
\begin{equation}
\sigma_{tot}(\omega )=\sigma_{el}(\omega )+\sigma_{in}(\omega ),
\end{equation}
where \begin{math} \sigma_{in}(\omega ) \end{math} is the inelastic
cross section.
In the commonly studied case in which the target ``absorbs'' radiation
one has a positive total cross section
\begin{math}\sigma_{tot} (\omega )>0\end{math}
since
\begin{equation}
\omega\, {\Im }m \left[\alpha (\omega +i0^+)\right]>0 \qquad {\rm
(absorption\  band)}.
\end{equation}
On the other hand, for a target in some excited energy state, (say)
with
atoms having ``inverted' energy populations, there will exist frequency
bands in which
\begin{equation}
\omega\, {\Im }m \left[\alpha (\omega +i0^+)\right]<0 \qquad {\rm
(amplifier\ band)}.\phantom{x}
\end{equation}
The optical theorem in the form of Eqs.(9) and (12) yield the
following.
\par \noindent
{\bf The Optical Klein Paradox:} If \begin{math}{\cal B}\end{math} is
the set of frequencies within which the target is an electromagnetic
amplifier, then
\begin{equation}
\sigma_{tot}(\omega \in {\cal B})<0.
\end{equation}

A negative total cross section in an amplifying frequency band is only
at first glance an impossibility. The purpose of this work is to
discuss physical meaning of the optical Klein Paradox for amplifiers
such as inverted population targets,
e.g. ``pumped'' LASER materials\cite{14}. For the case of the
Dirac equation, pair production allowed (without changing total charge)
for more electrons in an outgoing state than were present in the
incoming state. This simple fact makes transparent the notion of
a negative forward transmission coefficient \begin{math} P<0
\end{math}.
Similarly, for an excited state target one incident photon can give
rise
to two photons in the outgoing state when the excited state target is
induced to decay (say) into a target ground state. In an
electromagnetic
scattering experiment, there can be more radiation behind the target
than
that which would exist if the excited radiating target were removed.

In Sec.II, we consider an electromagnetic wave traveling through a
medium with a dielectric response function
\begin{math} \varepsilon(\zeta ) \end{math}.
For a dilute density per unit volume \begin{math} n \end{math} of
polarizable targets in the medium, the dielectric response
\begin{equation}
\varepsilon (\zeta )=1+4\pi n\alpha (\zeta )+\ ...\ \ {\rm if}\ \
|4\pi n\alpha (\zeta )| \ll 1,
\end{equation}
so that
\begin{equation}
{\Im}m \big[\varepsilon (\omega +i0^+)\big] = 4\pi n\,{\Im}m
\big[\alpha (\omega +i0^+)\big].
\end{equation}
The properties of
\begin{math}  \varepsilon (\omega +i0^+) \end{math} within an
amplifying
frequency band \begin{math} {\cal B} \end{math} will be explored.
In Sec.III, the concept of a negative noise temperature
\begin{math} T_n \end{math} will be defined. The definition of an
amplifying target will be related to the notion of negative noise
temperature \begin{math} T_n \end{math}~\cite{15,16}. It will
be shown that an amplifying frequency band
\begin{math} {\cal B} \end{math} may equally
well be defined by the notion of a negative noise temperature
\begin{math} T_n(\omega \in {\cal B})<0  \end{math}. The
electromagnetic
Klein paradox occurs in all material systems exhibiting a
negative radiation noise temperature. The general optical theorem will
be proved in Sec.IV without regard to the sign of the noise
temperature.
Physical examples will be discussed in the concluding Sec.V.

\section{A Traveling Plane Wave in Matter}

Consider an electromagnetic plane wave
\begin{equation}
{\bm E}={\Re}e\left[{\bm E}_0 e^{i(kz-\omega t)}\right]
\end{equation}
traveling through a medium with a dielectric response function
\begin{math} \varepsilon(\omega +i0^+) \end{math} so that
\begin{equation}
k=\frac{\omega }{c}\sqrt{\varepsilon(\omega +i0^+)}.
\end{equation}
The intensity of the light beam described by the plane wave
is proportional to
\begin{equation}
\overline{|{\bm E}|^2}=\frac{1}{2}|{\bm E}_0 |^2 \exp\left(-hz
\right)
\end{equation}
defines the extinction coefficient
\begin{equation}
h=2{\Im}m(k).
\end{equation}
From Eq.(18) it is evident that
\begin{equation}
\begin{matrix} {\rm an\ absorbing\ medium}\ \Longrightarrow \ (h>0)
\cr {\rm and} \cr {\rm an\ amplifying\ medium}\ \Longrightarrow \
(h<0). \end{matrix}
\end{equation}
For a medium consisting of a dilute gas of polarizable particles for
which \begin{math} n|\alpha (\omega +i0^+)| \ll 1 \end{math}, it
follows from Eqs.(14), (17) and (19) that
\begin{equation}
h=n\sigma_{tot}=4\pi n\left(\frac{\omega }{c}\right)
{\Im }m\alpha (\omega +i0^+).
\end{equation}
Eqs.(20) and (21) imply the optical Klein theorem:
\begin{equation}
\begin{matrix} {\rm an\ absorbing\ medium}\ \Longrightarrow \
(\sigma_{tot}>0) \cr {\rm and} \cr {\rm an\ amplifying\ medium}\
\Longrightarrow \ (\sigma_{tot}<0).
\end{matrix}
\end{equation}
The paradox of having
\begin{math} \sigma_{tot}(\omega \in {\cal B})<0 \end{math}
in an amplifying frequency band \begin{math} {\cal B} \end{math}
has now been formally proved.

\section{Noise Temperature and Spectral Functions}

The quantum spectral functions for dipole moment
\begin{math} {\bm p}(t) \end{math} fluctuations in a target
may be defined by
\begin{equation}
S_\pm (\omega )= \frac{1}{3}\sum_{I,F}
p_I \, \left|\left<F\left|{\bm p}\right|I \right>\right|^2 \delta
\left[\omega \mp \frac{(E_F-E_I)}{\hbar}\right],
\end{equation}
where \begin{math} p_I \end{math} is the target probability
of being in an initial energy eigenstate
\begin{math} H\left|I \right >=E_I\left|I \right> \end{math}.
Defining the free energy \begin{math} {\cal F}  \end{math}
so that the initial probability is normaized to
\begin{math} \sum_I p_I=1   \end{math},
if the initial target is at temperature \begin{math} T \end{math},
\begin{equation}
p_I{\rm (Thermal)}=
\exp\left(\frac{{\cal F}-E_I}{ k_B T}\right)
\ \ {\rm (Equilibrium)},
\end{equation}
then the spectral functions in Eq.(23) obey the detailed balance
condition
\begin{equation}
S_-(\omega )=S_+(\omega )\exp\left(-\frac{\hbar \omega }{k_BT}\right)
\ \ {\rm (Equilibrium)}.
\end{equation}
The notion of having a noise temperature
\begin{math} T_n(\omega ) \end{math}
which depends on frequency follows from the definition
\begin{equation}
S_-(\omega )=S_+(\omega )
\exp\left[-\frac{\hbar \omega }{k_BT_n(\omega )}\right]
\ \ {\rm (General)}.
\end{equation}

The Kubo formula for the electric polarizability of the target,
\begin{equation}
\alpha (\zeta )=\left(\frac{i}{3\hbar }\right)\int_0^\infty e^{i\zeta
t} \left<{\bm p}(t)\cdot{\bm p}(0)-{\bm p}(0)\cdot{\bm
p}(t)\right>dt,
\end{equation}
may be written in terms of the spectral functions in Eq.(23)
employing
\begin{equation}
\left<{\bm p}(t)\cdot{\bm p}(0)- {\bm p}(0)\cdot{\bm p}(t)\right> =
3\int_{-\infty}^\infty e^{-i\omega t} \left[S_+(\omega )-S_-(\omega
)\right]d\omega .
\end{equation}
Eqs.(27) and (28) imply
\begin{equation}
 \alpha (\zeta )= \frac{1}{\hbar}\, \int_{-\infty }^\infty
\left[\frac{S_+(\omega )-S_-(\omega )}{\omega -\zeta }\right] d\omega ,
\end{equation}
from which one may deduce that
\begin{equation}
{\Im }m\left(\alpha (\omega +i0^+)\right)=
\left(\frac{\pi}{\hbar}\right)\left[S_+(\omega )-S_-(\omega ) \right].
\end{equation}
Eqs.(9), (26) and (30) imply
\begin{equation}
\sigma_{tot}(\omega )=\left(\frac{4\pi^2 \omega }{\hbar c}\right)
\left[1-e^{-\hbar \omega /k_BT_n(\omega )}\right]S_+(\omega ).
\end{equation}
One may define the ``symmetrical'' noise spectral function
according to \begin{math}\bar{S} =(1/2)(S_+ +S_-)\end{math}.
In terms of
\begin{equation}
\bar{S}(\omega )=\frac{1}{2} \left[1+e^{-\hbar \omega / k_BT_n(\omega
)}\right]S_+(\omega ),
\end{equation}
which in virtue of Eq.(23) obeys
\begin{equation}
\bar{S}(\omega )\ge 0,
\end{equation}
Eq.(31) reads
\begin{equation}
\sigma_{tot}(\omega )=\left(\frac{8\pi^2 \omega }{\hbar c}\right)
\tanh\left[\frac{\hbar \omega }{k_B T_n (\omega )}\right]
\bar{S}(\omega ).
\end{equation}

Eqs.(33) and (34) are the central results of this section. They
allow Eq.(22) to be written in terms of the noise temperature as
\begin{equation}
\begin{matrix} ({\rm absorbing})\ T_n(\omega )>0 \ \Longrightarrow \
(\sigma_{tot}(\omega )>0) \cr {\rm and} \cr ({\rm amplifying})\
T_n(\omega )<0 \ \Longrightarrow \ (\sigma_{tot}(\omega )<0).
\end{matrix}
\end{equation}
Thus, the frequency band \begin{math} {\cal B} \end{math} in Eq.(13)
which is amplifying can be characterized as having a negative noise
temperature \begin{math} T_n(\omega \in {\cal B})<0 \end{math}.
As will be discussed in the concluding Sec.V, a variety of
negative temperature systems\cite{15,16} can be thought to exhibit the
electromagnetic Klein paradox.

\section{Generalized Optical Theorem}

In this section it is shown why the optical theorem of
Eq.(8) holds true for an elastic scattering amplitude at
positive or negative noise temperature. The negative noise
temperature regime gives rise to a negative total cross section.

Suppose an electromagnetic wave is incident on a target localized near
the origin of the spatial coordinate system. The electric field far
from the target then has the familiar scattering form as
\begin{math} r\to \infty  \end{math}; i.e.
\begin{equation}
{\bm E}\to {\Re}e \left\{E_0 e^{-i\omega t} \left({\bm e}_i
e^{i\omega z/c}+ {\bm F\cdot e}_i\frac{e^{i\omega r/c}}{r}\right)
\right\},
\end{equation}
where \begin{math} {\bm F} \end{math} is the dyadic elastic
scattering amplitude
\begin{equation}
F_{i\to f}={\bm e}^*_f {\cdot \bm F \cdot}{\bm e}_i .
\end{equation}
The intensity of the incident wave in Eq.(36) is given by
\begin{equation}
I_0=\left(\frac{c|E_0|^2}{8\pi }\right).
\end{equation}
The intensity of the full wave is given by
\begin{equation}
I=\left(\frac{c\overline{|{\bm E}|^2}}{4\pi}\right).
\end{equation}

The optical theorem proof envisages detecting radiation
directly behind the target but far away. We employ polar
coordinates
\begin{math} {\bm r}=(x,y,z)=({\bm r}_\perp ,z) \end{math}
and the limits \begin{math} z\to \infty \end{math} so that
\begin{math} r_\perp \ll z \end{math}. The total cross section
is then usually viewed in terms of the ``missing intensity''
detected on a screen placed directly behind but far away from the
target. In mathematical terms, the total cross section may be
defined as
\begin{equation}
\sigma_{total}=\lim_{z\to \infty }\int
\left[\frac{I_0-I({\bm r}_\perp,z) }{I_0}\right]\,d^2{\bm r}_\perp .
\end{equation}

It is crucial to understand the physical meaning of Eq.(40). If the
intensity behind the screen \begin{math} I  \end{math} is less than
the incident intensity \begin{math} I_0 \end{math}, i.e. if
\begin{math} I({\bm r}_\perp,z\to \infty )<I_0 \end{math}, then the
total cross section is positive
\begin{math} \sigma_{tot}>0 \end{math} which is most often
presumed. The intensity behind the target is usually less
than in the incident radiation because
(i) the radiation is elastically scattered away at some angle
with cross section \begin{math} \sigma_{el} \end{math} or
(ii) the radiation was inelastically absorbed by the target
with cross section \begin{math} \sigma_{in} \end{math}.
Altogether,
\begin{math} \sigma_{tot}=\sigma_{el}+\sigma_{in} \end{math}
as in Eq.(10). But now consider Eq.(40) if the target is in an
excited state. The target can be induced to decay by the incident
radiation making the intensity on the behind the target screen
bigger than the intensity of the incident radiation; i.e.
\begin{math} I({\bm r}_\perp,z\to \infty )>I_0 \end{math} is a
distinct possibility due to the {\em added radiation} from the
decaying target. The excited target case
\begin{math} \sigma_{tot}<0 \end{math} is thus quite possible.

In either of the above cases the optical theorem of Eq.(8) holds
true as we shall now prove. From Eqs.(36)-(39) it follows that
as \begin{math} z\to \infty \end{math} and as
\begin{equation}
r=\sqrt{z^2+r_\perp ^2}\to z+(r_\perp ^2/2z)+ ...\ ,
\end{equation}
the intensity on the screen behind the target obeys
\begin{equation}
\left[\frac{I_0-I({\bm r}_\perp,z) }{I_0}\right]\to -{\Re
}e\left[\frac{2F_{i\to i}e^{i(\omega r_\perp ^2/2cz)}}{z}\right],
\end{equation}
which comes from the interference term when one takes the
absolute square of the amplitude in Eq.(36). Now,
Eqs.(40) and (42) imply the total cross section
\begin{equation}
\sigma_{tot}=-\lim_{z\to \infty}{\Re }e\left\{\int
\left[\frac{2F_{i\to i}e^{i(\omega r_\perp ^2/2cz)} }{z}\right]\,
d^2{\bm r}_\perp
\right\}.
\end{equation}
Using
\begin{math}
\int (...) d^2{\bm r}_\perp \to 2\pi \int_0^\infty (...)r_\perp
dr_\perp
\end{math}
allows for a simple evaluation of the integral in Eq.(43) leading
to the optical theorem
\begin{equation}
\sigma_{tot}=
\left(\frac{4\pi c}{\omega }\right){\Im}m \big(F_{i\to i}\big).
\end{equation}

It is important to note that the proof of the optical theorem
given above in no manner invokes the sign of the total cross section.
If the net intensity on the detectors directly behind the target
is less than the incident intensity, then
\begin{math} \sigma_{tot}>0 \end{math}.
If the net intensity on the detectors directly behind the target
is more than  the incident intensity, then
\begin{math} \sigma_{tot}<0 \end{math}. In either case Eq.(44) holds
true.

\section{Conclusions}

We have shown that the electromagnetic Klein paradox arises
in the form of a possible negative total cross section
\begin{math} \sigma_{tot} \end{math} for radiation to scatter
off a target in an excited state. An excited state target can
be viewed as having a negative noise temperature. Negative noise
temperature targets are a reality\cite{17,18,19} for laser or
maser pumped materials.

Some of these ideas may be of basic interest and should be explored
in detail for the strong interaction part of the standard model
which invokes quantum chromodynamics (QCD) of quarks and gluons. At
sufficiently small length scales, the effective interactions between
the color charged particles within the plasma are thought to be
weak. Short distance asymptotic freedom provides the perturbation
theory basis for comparisons between QCD theory and scattering
experiments. We have previously shown\cite{31,32,33} that the
asymptotically free vacuum has ``negative dissipation'' in the color
dielectric response function $\epsilon (Q^2)$. The color dielectric
response function determines the strong interaction coupling
strength via $\alpha_{QCD}(Q^2)= g^2/\{4\pi \hbar c
\epsilon_{QCD}(Q^2)\}$. Negative dissipation implies an
asymptotically free QCD excited state amplifier unstable to decay.

The fact that the Landau ghost does not haunt QCD perturbation
theory is totally dependent upon the existence of negative
color dissipation. It has been shown in our earlier work that
 while asymptotic freedom exorcises the Landau ghost from QCD,
  it ladens it with negative dissipation. A direct consequence
   of confinement and asymptotic freedom  in QCD is the notion
   that an applied color electric field produces
   an ``anti-Ohm's law'' color current which then  cools the
   perturbative QCD vacuum.

Physical negative dissipation can occur when matter is pumped
up by an external energy source.
The resulting ``amplifier'' excited state can then return the
 energy when the matter decays back into the true ground state.

The physical interpretation of vacuum cooling induced by the
application of a high frequency chromo-dynamic field is (at first
glance) more than a little obscure. How might one cool what is
already asserted to be the vacuum? A true quantum vacuum is
conventionally visualized as having the lowest possible energy. No
further cooling should then be possible. Any notion of asymptotic
freedom as a justification of QCD perturbation theory must then
refer to an excited state color current amplifier with negative
resistance. One may indeed cool down an excited state. We have
then been forced to conclude that the asymptotically free QCD
vacuum is in reality unstable which must eventually decay into the
true vacuum. But the situation with regard to comparisons between
theory and experiment needs further elucidation.

Experimental achievement of negative noise temperatures,
characteristic of electronic energy levels with inverted populations,
was absolutely essential for the construction of the original maser
and laser devices. Negative noise temperatures also occur naturally
within some matter clouding galaxies. The negative dissipative part
of the electrical conductivity in some astrophysical matter is made
manifest via the measured negative absorption (i.e. amplification)
of the electromagnetic radiation that passes through these materials.
However, we must also realize that amplification in excited state
systems can only have a finite lifetime. The excited state atoms
decay into their ground states after giving rise to a burst of
electromagnetic radiation. To revert again to the excited state
amplifier status, energy must be pumped back into the matter by
an external source. The instability of the QCD vacuum needs to
be investigated in depth in view of the results of the present paper.

A negative electromagnetic cross section simply means there is
more radiation energy behind the target in the outgoing state
than there would be if the target were {\em not} present.
The target then represents an electromagnetic amplifier.
The amplifier must be pumped into an excited state by an external
energy source before the next incident wave is sent to scatter off
the target. Such amplifiers have been of considerable interest in
astrophysics\cite{20,21,22,23,24} where they occur naturally
in interstellar media.

The relativistic Klein paradox does not in reality imply any sort of
negative probability. In the Dirac theory, one may create (from pair
production) more electrons in the final state than were present in
the initial state. But in probability terms the {\em total charge}
would remain conserved\cite{25,26,27,28,29,30}. Similarly, in the
Maxwell theory, one may create (from atomic quantum state decays)
more photons in the final state than were present in the initial
state. But in probability terms the {\em total energy} would remain
conserved. From the above optical theorem proof, the total cross
section for the scattering of electromagnetic radiation may obey
\begin{math} \sigma_{tot}(\omega \in {\cal B})<0  \end{math}
because the radiation energy in the final state may exceed (by
far) the radiation in the initial state, but the probability for
the scattering event is nevertheless positive. In this manner,
the electromagnetic Klein paradox may be resolved.

\end{document}